\documentclass[10pt,conference]{IEEEtran}
\IEEEoverridecommandlockouts
\usepackage{cite}
\usepackage{hyperref}
\usepackage{amsmath,amssymb,amsfonts}
\usepackage{algorithmic}
\usepackage{graphicx}
\usepackage{textcomp}
\usepackage[table]{xcolor}

\usepackage[utf8]{inputenc}
\usepackage[american]{babel}
\usepackage{amsmath}
\usepackage{amssymb}
\usepackage{array}
\usepackage{csquotes}

\usepackage{amsthm}
\usepackage{multirow}
\usepackage{algorithm}
\usepackage{fixltx2e}
\usepackage{amsthm}
\usepackage[inline]{enumitem}
\usepackage{caption}
\usepackage{todonotes}

\usepackage{tikz}

\usepackage{url}

\usepackage{xspace}
\newcommand{\cf}[0]{\textit{cf.}\@\xspace}

\newcommand{\eg}[0]{\textit{e.\,g.},\@\xspace}

\newcommand{\etal}[0]{\textit{et\,al.}\@\xspace}
\newcommand{\sect}[1]{Section\xspace\ref{sec:#1}}

\newcommand{\fig}[1]{\figurename\xspace\ref{fig:#1}}

\newcommand{\tab}[1]{Table\xspace\ref{tab:#1}}

\newcommand{\goal}[1]{\textbf{G#1}}
\newcommand{\rqs}[1]{\textbf{RQ#1}}

\def\BibTeX{{\rm B\kern-.05em{\sc i\kern-.025em b}\kern-.08em
    T\kern-.1667em\lower.7ex\hbox{E}\kern-.125emX}}
\usepackage{tcolorbox}
\newtcolorbox{mybox}{colback=!5!lightgray,colframe=black}

\begin{document}
\title{A User-Study Protocol for Evaluation of Formal Verification Results and their Explanation}

\author{\IEEEauthorblockN{Arut Prakash Kaleeswaran, Arne Nordmann}
\IEEEauthorblockA{\textit{Corporate Sector Research,} \\
\textit{Robert Bosch GmbH}\\
  Renningen, Germany \\
\{arutprakash.kaleeswaran, arne.nordmann\}@de.bosch.com}
\and
\IEEEauthorblockN{Thomas Vogel, Lars Grunske}
\IEEEauthorblockA{\textit{Software Engineering Group,}\\
\textit{Humboldt-Universit\"{a}t zu Berlin}\\
 Berlin, Germany \\
\{thomas.vogel, grunske\}@informatik.hu-berlin.de}
}
\maketitle
\begin{abstract}
\textit{Context:} Ensuring safety for any sophisticated system is getting more complex due to the rising number of features and functionalities.
This calls for formal methods to entrust confidence in such systems. Nevertheless, using formal methods in industry is demanding because of their lack of usability, \eg the difficulty of understanding verification results.
Thus, our hypothesis is that presenting verification results of model checkers in a user-friendly manner could promote the use of formal methods in industry.
\textit{Objective:} We aim to evaluate the acceptance of formal methods by Bosch automotive engineers, particularly whether the difficulty of understanding verification results can be reduced.
\textit{Method:} We perform two different exploratory studies. First, we conduct an online survey to explore challenges in identifying inconsistent specifications and using formal methods by Bosch automotive engineers. Second, we perform a one-group pretest-posttest experiment to collect impressions from Bosch engineers familiar with formal methods to evaluate whether understanding verification results is eased by our counterexample explanation approach.
\textit{Limitations:} The main limitation of this study is its generalization, since the survey focuses on a particular target group and uses a pre-experimental design. 
\end{abstract}


\section{Introduction}
\label{sec:introduction}

\begin{figure}[!tbh]
	\centering
	\includegraphics[width=\linewidth]{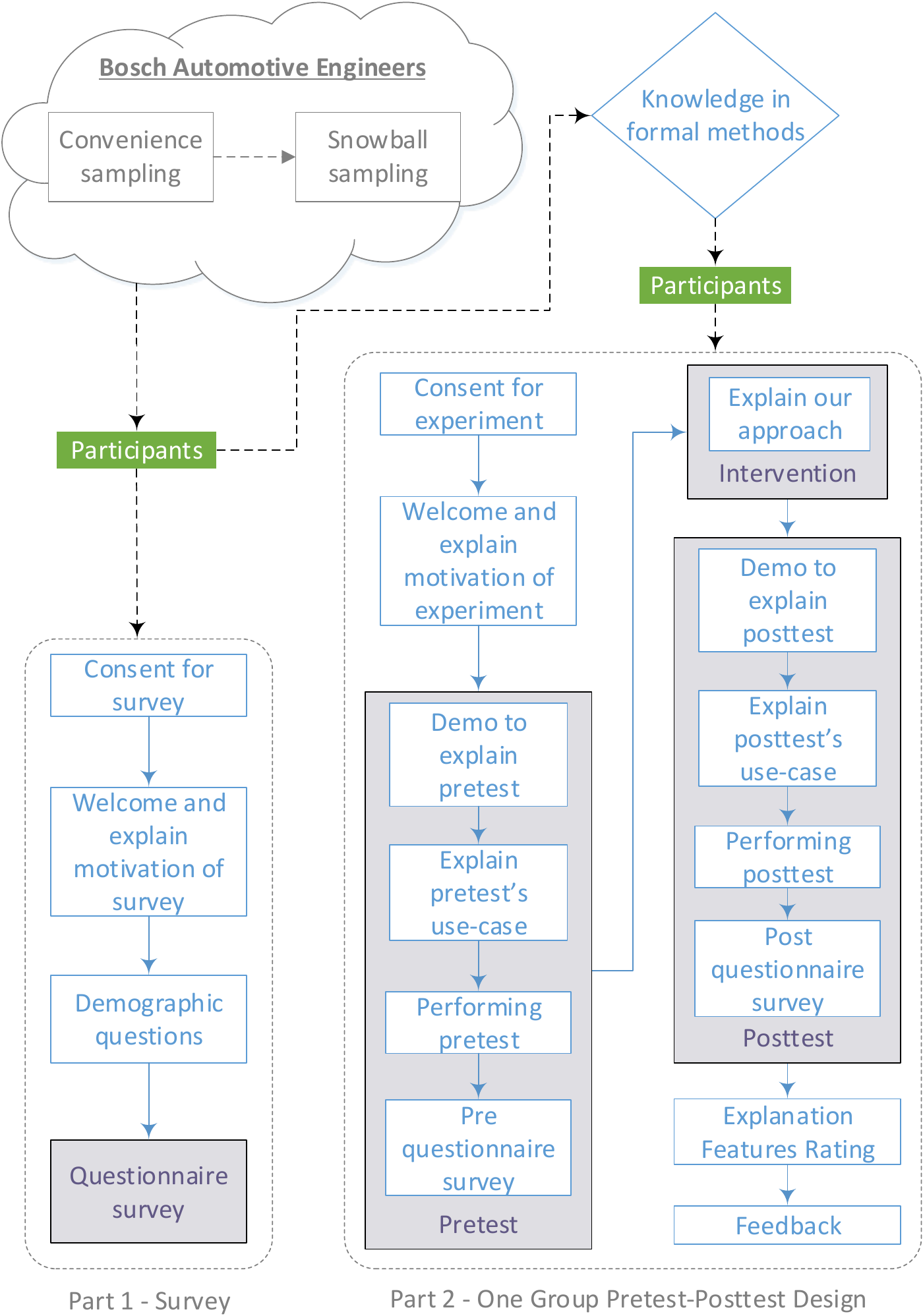}
	\caption{ \small Overview of the proposed study. \emph{Part\,1} is an online survey performed with a wide range of participants, \emph{Part\,2} is a one-group pretest-posttest experiment performed with formal methods experts. Gray color boxes indicate the main tasks, \eg survey questionnaire and pretest and posttest experiments.}
	\label{fig:steps}
	\vspace{-1.5em}
\end{figure} 

During the development of safety-critical systems, as and when the requirements or the system specification change, the consistency of the system specification must be verified. In an industrial setting where this re-verification is done almost always manually~\cite{WaliaC09}, contract-based design (CBD)~\cite{CimattiT12} can substitute this manual work by automating the verification process using a model checker (\cf~Fig.\xspace1 and Sect.\xspace2 in~\cite{KaleeswaranNVG20} for an example).
Whenever an inconsistency is found during the verification, the model checker exemplifies the violation by generating a counterexample. It is then up to an engineer to understand the counterexample and to identify the root cause of the violation by manually tracing the violation back from the counterexample to the  original system specification.

Identifying the inconsistent specifications from a set of specifications is challenging, though, because specifications of real-world use cases can comprise hundreds of pages~\cite{Schuppan16}.
Further, identifying faults from a counterexample is error-prone and time-consuming, especially for non-experts in formal methods because counterexamples are lengthy and cryptic~\cite{BergSJ07,LeueB12,MuramTZ15,BarbonLS19,OvsiannikovaBPV21}. 
Thus, an automated method for explaining counterexamples is highly desirable to assist engineers in understanding counterexamples and thus, in identifying faults in their models.

Formal methods are not new to Bosch. They are used to specify requirements as pattern-based specifications to support verification during product development~\cite{PostMHP12,PostH12}. 
Additionally, we have presented a \emph{counterexample explanation approach} that attempts to ease the use of formal methods by reducing the manual work and difficulty of interpreting the verification results generated by model checkers~\cite{KaleeswaranNVG20}. 
Particularly, we target refinements in system design and the verification of their consistency. Usually, engineers refine a top-level component and its specification into sub-components and their respective individual specifications. A model checker can verify the consistency of such a refinement. 
If an inconsistency was introduced during the refinement by an engineer, the model checker returns a counterexample. Our approach provides an additional explanation of this counterexample to engineers in order to ease understanding of the model checker result and identifying the inconsistency of the refinement.

To explore whether our counterexample explanation approach~\cite{KaleeswaranNVG20} does improve the understanding of the model checking results, we intend to perform a \emph{one-group pretest-posttest experiment}. Since we want to perform the study with professional engineers working at Bosch, our study requires working time from these engineers and thus implicitly incurs costs. Thus, the number of engineers to participate in the study will be limited. The one-group pretest-posttest experiment supports conducting the study with a limited number of participants. Further, to explore the general acceptance of formal methods and to contemplate on challenges and the complexity faced by engineers in identifying inconsistent specifications, we intend to perform an \emph{online survey}. In this paper, we summarize the research questions that we evaluate, the design and execution plan of the study, target participants, analysis plan, and threats to validity.

\section{Research Questions}
\label{sec:rq}

Our study aims to explore and understand the challenges in identifying inconsistent specifications, and the acceptance of formal methods by Bosch automotive engineers. Therefore, this user study has two significant goals: \textbf{(G1)}~to understand challenges faced by Bosch engineers in order to identify inconsistent specifications and challenges along with their opinions to use formal verification or formal methods in real-world development processes, and \textbf{(G2)}~to explore whether Bosch engineers are interested in using formal methods, particularly model checking, in real-world development processes if the difficulty of understanding model checking results is reduced by our counterexample explanation approach. Considering these two goals, we formulate the following research questions:

\noindent\emph{\rqs{1}: To what extent do engineers face challenges in identifying inconsistent specifications in formal models that are introduced during a refinement of a system?}\\
\noindent With this RQ we want to investigate whether:
\begin{itemize}[leftmargin=*]
	\item Understanding formal notations is difficult for engineers.
	\item Identifying inconsistent specifications that are introduced during a refinement of a top-level specification is difficult.
\end{itemize}

\noindent\emph{\rqs{2}: To what extent is identifying inconsistent specifications and using formal methods beneficial to a real-world development process?}
\\
\noindent With this RQ we want to investigate whether:
\begin{itemize}[leftmargin=*]
\item Usage of formal verification or formal methods is beneficial in a real-world development process.
\item Identifying inconsistent specifications is beneficial in a real-world development process.
\end{itemize}

\noindent\emph{\rqs{3}: To what extent do engineers prefer to use formal methods (model checkers particularly) if the difficulty is reduced for understanding verification results to identify inconsistent specifications?}
\\
\noindent With this RQ we want to investigate whether:
\begin{itemize}[leftmargin=*]
\item The counterexample explanation approach eases comprehension compared to interpreting the raw model checker output for engineers with a formal methods background.
\item The counterexample explanation approach is understandable by engineers with a background in formal methods.
\item It is possible for engineers with a background in formal methods to identify and fix inconsistent specifications based on the counterexample explanation approach.
\item The counterexample explanation approach can promote formal verification and usage of model checking in real-world development processes.	
\end{itemize}

\section{Variables}
\label{sec:variables}

To attain the goals \textbf{G1} and \textbf{G2}, we perform two different types of exploratory user studies as shown in \fig{steps}. The first study is an \emph{online survey (Part\,1)}, the second study is a \emph{one-group pretest-posttest user study (Part\,2)}. 

\subsection{Variables of Part\,1: Online Survey}
\label{sec:var_part1}

Our online survey evaluates the research question \rqs{1} and \rqs{2}. The independent variables of \emph{Part\,1} are \emph{participants' professional background and experience}. The dependent variables are different for each research questions. For the research question  \rqs{1}, the dependent variable is the \emph{difficulty in understanding} that infers that understanding formal notations and identifying inconsistent specifications by engineers are difficult. Similarly, the dependent variable for \rqs{2} is the \emph{increase in confidence in system safety}, that is, the identification of inconsistent specifications and use of formal methods in real-world development processes can make systems safer.

\subsection{Variables of Part\,2: One-Group Pretest-Posttest Design}
\label{sec:var_part2}

As per Babbie~\cite{Babbie16}, an experimental stimulus (also called an intervention) is the independent variable. In the one-group pretest-posttest design, we use our counterexample explanation approach as the intervention. Therefore, it serves as the independent variable of \emph{Part\,2}. Further, the research question \rqs{3} are evaluated based on the following four attributes that serve as dependent variables for \emph{Part\,2} of our study:

\begin{enumerate}[leftmargin=*]
	\item \emph{Better understanding:} Does the counterexample explanation approach allow engineers to understand model checking results and identify inconsistencies more effectively?
	\item \emph{Quicker understanding:} Does the counterexample explanation approach allow engineers to understand model checking results and identify inconsistencies more efficiently?
	\item \emph{Confidence:} Does the counterexample explanation approach make engineers more confident in their understanding of the system and its inconsistency resp. safety?
	\item \emph{No value:} This attribute is inversely related to the above attributes. Will the counterexample explanation approach provide no or only minimal value to real-world projects?
\end{enumerate}

\section{Design of the User Study}
\label{sec:design}

In this section, we describe the design, questionnaires, and tools used for both the \emph{online survey (Part\,1)}, and the \emph{one-group pretest-posttest user study (Part\,2)}.

\subsection{Part\,1: Online Survey}
\label{sec:part1}

For \emph{Part\,1}, we use a cross sectional survey~\cite{KitchenhamP08} to collect data from engineers to achieve the objective of \goal{1}. For planning and conducting this online survey, we follow the guidelines of Neuman~\cite[Chapter~7]{Neuman14} (majorly), Kitchenham and Pfleeger~\cite{KitchenhamP08}, and Fink~\cite{Fink03}. In addition to Neuman~\cite{Neuman14}, we follow Robson and McCartan ~\cite[Chapter~11]{RobsonM16}, and Babbie [5, Chapter 9] for the questionnaire construction. Further, we refer to and adapt some of the questionnaires from existing user surveys by Gleirscher and Marmsoler~\cite{GleirscherM20}, and Garavel \etal~\cite{GaravelBP20}. Gleirscher and Marmsoler~\cite{GleirscherM20} perform the largest cross sectional survey with 216 participants to study the existing and intended use of formal methods. Similarly, Garavel \etal~\cite{GaravelBP20} conduct a user survey with 130 participants and 30 questions to collect information on the past, present, and future of formal methods in research, industry, and education. Our main contribution wrt. similar surveys is: (1) we particularly focus on identifying challenges that engineers face in identifying inconsistent specifications, not general challenges of using formal methods, and (2) the study is performed with engineers who work on real-world automotive projects.

\begin{table*}[bth]
	\centering
	\caption{Online survey questions for study Part\,1.}
	\label{tab:survey_questions}
	\begin{tabular}{p{0.55cm}|m{11cm}|l|m{2cm}}
		\hline \rowcolor{gray!10}
		\textbf{Label} & \textbf{Questions}                                                                                                            & \textbf{Scale}        & \textbf{Label (\cf \tab{likert_scale})} \\ \hline \rowcolor{gray!10}
		\multicolumn{4}{l}{\textbf{Demographic Questions}}                                                                                                                                                   \\ \hline
		Q1             & Rate your knowledge of formal methods                                                                                              & Ordinal &            LS1                   \\ \hline
		Q2             & How many years have you used formal methods in your daily work? (answer the duration separately for academia and industry)                                             & Ordinal &       LS2                        \\ \hline
		Q3             & List the application(s) you worked on using formal methods (if available)                                                                & Nominal               &     Not Applicable                          \\ \hline
		Q4             & How many years have you worked in the safety domain? (only in industries)                                                                        & Ordinal &         LS2                      \\ \hline
		Q5             &  List the application(s) you worked on focusing on safety aspects (if available)                                                                & Nominal               &       Not Applicable                        \\ \hline  \rowcolor{gray!10}
		\multicolumn{4}{l}{\textbf{Main Survey Questions}}                                                                                                                                                   \\ \hline
		Q6             & How easy it is for you to identify inconsistent specifications?                                                                       & Nominal   and Ordinal &                LS3               \\ \hline

		Q7             & What sort of methods do you use to identify inconsistent specifications?                                                 & Nominal &     Not Applicable                          \\ \hline
		Q8             & How fast could you identify inconsistent specifications?                                                                      & Nominal   and Ordinal &                  LS4             \\ \hline
		Q9            & What are the challenges that you face in order to identify inconsistent specifications?                                               & Nominal               &      Not Applicable                         \\ \hline
		Q10            & How easy is it for you to maintain consistency when refining requirements   for sub-components?                                         & Nominal   and Ordinal &   LS3                            \\ \hline
		Q11            & In your opinion, how beneficial is the identification of inconsistent specifications for a safety analysis?                                          & Nominal   and Ordinal &      LS5                         \\ \hline
	    Q12            & How hard is it for you to check the consistency of requirements that are   associated with components?                                & Nominal   and Ordinal &     LS3                          \\ \hline
		Q13            & How easy is it for you to understand formal notations?                                                                        & Nominal   and Ordinal &             LS3                  \\ \hline
		Q14            & What is your opinion on using formal verification?                                                                            & Nominal               &     Not Applicable                          \\ \hline
		Q15            & In your opinion, will the usage of formal verification make systems safer?                                                                      & Nominal   and Ordinal &                   LS5            \\ \hline
		Q16            & In your opinion, can the formal verification be an add-on to the functional safety methods   to ensure safety?                                         & Nominal   and Ordinal &      LS5                         \\ \hline
		Q17            & Can you imagine using formal methods if their understanding of formal notations is made easier? & Nominal   and Ordinal &                   LS5           \\ \hline
		Q18            & Do you think formal methods are usable in real-world development processes?                                              & Nominal   and Ordinal &      LS5                         \\ \hline
	\end{tabular}
\end{table*}

\tab{survey_questions} presents the questionnaire prepared for our online survey.  
The response for each question is captured either as qualitative statements, a set of predefined scale answers, or a combination of both. We use a 7-point scale as it increases the reliability of answers from participants over a 5-point scale according to Joshi \etal~\cite{JoshiKCP15}. The scale answers set we use in this survey is listed in \tab{likert_scale}.

\begin{table*}[!tbh]
	\centering
	\caption{Likert and answer scales used for this study.}
	\label{tab:likert_scale}
	\begin{tabular}{p{0.5cm}|m{1.2cm}|m{14.7cm}}
		\hline
		\rowcolor{gray!10}
		\textbf{Label} & \textbf{Type} & \textbf{Likert \& Answer Scales}                                                                                                                                                               \\ \hline
		LS1                                   & Expertise                          & Novice,  Advanced Beginner, Competent, Proficient,   Expert, Mastery, Practical Wisdom, No Opinion                                                                                                   \\ \hline
		LS2                                   &  Experience                 & $>=$\,1, $<$\,1 to 2, $<$\,2 to 4, $<$\,4 to 6, $<$\,6 to 8, $<$\,8 to 10, $<$\,10, No   Experience                                                                                                                      \\ \hline
		LS3                                   & Agreement                          & Extremely Hard, Hard, Slightly   Hard, Neither Hard nor Easy, Slightly Easy, Easy, Extremely Easy, No Opinion                                                                                                                \\ \hline
		LS4                                   & Agreement                          & Extremely Fast, Fast, Slightly   Fast, Neither Fast nor Slow, Slightly Slow, Slow, Extremely Slow, No Opinion                                                                                                                \\ \hline	
		LS5                                   & Likelihood                         & Definitely,  Very Probably,  Probably, Neither Probably or Possibly,   Possibly,  Probably Not,  Definitely Not, No Opinion                                                                                                       \\ \hline
		LS6                                   & Agreement                         & Strongly agree,  Agree,  Somewhat agree,    Neither agree or disagree, Somewhat disagree,  Disagree, Strongly disagree, No Opinion                                                                                                       \\ \hline
     	LS7                                   & Usefulness                         & Exceptional, Excellent, Very Good, Good, Fair, Poor, Very Poor, No Opinion                                                                                                       \\ \hline
	\end{tabular}
\end{table*}

\subsection{Part\,2: One-Group Pretest-Posttest Design}
\label{sec:part2}

\emph{Part\,2} of our study is an exploratory pre-experimental user study following a \emph{one-group pretest-posttest design} to attain goal \goal{2}. We follow the guidelines by Campbell and Stanley~\cite{CampbellS63} to conduct this part of our study.

One of the main drawbacks of using a one-group pretest-posttest design is that it does not meet the scientific standards of an experimental design. For example, the pre-experimental study designs does not have a control group like a true experiment~\cite{WohlinRHO12}. Thus, comparison and generalization of the results based on the provided intervention/stimulus may not be possible. However, we intend to use this pre-experimental user study design because of the scarcity of participants. To find a considerable number of participants (30 to 40) with knowledge of formal methods and model checkers inside an industrial organization is ambitious. Performing a true experiment with a lower number of participants raises the threat to external validity.
Therefore, we intend to perform a one-group pretest-posttest experiment with Bosch automotive engineers that allows us to capture results from real-world user behavior, even with a limited number of participants. However, the pre-experimental study has several internal and external threats to be considered. We discuss handling of the threats listed by Campbell and Stanley~\cite[Table 1]{CampbellS63} in \sect{tov}.

Along with the guidelines by Campbell and Stanley, we refer to the protocol by Zaidman \etal~\cite{ZaidmanMSD13} for a one-group pretest-posttest experiment. They evaluate a tool called \emph{FireDetective} that supports understanding of Ajax applications at both the client-side (browser) and server-side. Their evaluation is performed using two user study variants (i)~pretest-posttest user study, and (ii)~a field user study, where the former is performed with eight participants and the latter is performed with two participants. We plan to perform the one-group pretest-posttest experiment with Bosch automotive engineers and discard the field user study for our evaluation.

The questionnaire presented in \tab{onegroup} is used for the one-group pretest-posttest study (\emph{Part\,2} of our overall study). Similar to \emph{Part\,1}, the response for each question is either a qualitative statement, or a set of predefined scale answers with 7-point scale, or a combination of both.

\begin{table*}[!tbh]
	\centering
	\caption{One-group pretest and posttest questions.}
	\label{tab:onegroup}
	\begin{tabular}{p{0.5cm}|m{11.5cm}|l|m{2cm}}
		\hline \rowcolor{gray!10}
		\textbf{Label} & \textbf{Questions}                                                                                                            & \textbf{Scale}        & \textbf{Label (\cf \tab{likert_scale})} \\ \hline \rowcolor{gray!10}
		\multicolumn{4}{l}{\textbf Task Questions}                                                                                                                                                                                                                                            \\ \hline
		TQ1                                  & How difficult was this use case for you to understand?                                                                                                    & Nominal and Ordinal                 & LS5                                                                        \\ \hline
		TQ2                                  & Do  you think this use case is difficult?                                                                                                                     & Nominal and Ordinal                 & LS5                                                                        \\ \hline
		TQ3                                  & Do  you think you have understood  results   from the model checker? (This question is only for pretest)                                          & Nominal and Ordinal                 & LS5                                                                        \\ \hline
		TQ4                                  & Do   you think you have understood the    explanations? (This question is only for the posttest)                                                       & Nominal and Ordinal                 & LS5                                                                        \\ \hline
		TQ5                                  & Of the following list, please select the inconsistent components.       & Nominal                             & Not Applicable                                                             \\ \hline
		TQ6                                  & Of the following list, please select the inconsistent specifications.                                                                                                               & Nominal                             & Not Applicable                                                             \\ \hline
		TQ7                                  & Please explain the reason that makes the specifications inconsistent from your understanding.                                                                                                                     & Nominal                             & Not Applicable                                                             \\ \hline		TQ8                                  & Please provide a solution to fix the inconsistency from your understanding.                                                                                                       & Nominal                             & Not Applicable                                                             \\ \hline
		TQ9                                  & Please provide a nominal behavior that is expected in the counterexample's erroneous states from your understanding.                                                       & Nominal                             & Not Applicable                                                             \\ \hline \rowcolor{gray!10}
		\multicolumn{4}{l}{\textbf{Understanding Model   Checker Results (Pretest)}}                                                                                                                                                                                                                                           \\ \hline
		PRQ1                                 & The results from the model checker allow me to understand the inconsistencies.                                                                    & Nominal and Ordinal                             & LS6                                                                        \\ \hline
		PRQ2                                 & The results from the model checker make me confident that I really understand the inconsistencies that I am investigating.              & Nominal and Ordinal                             & LS6                                                                        \\ \hline
		PRQ3                                 & The value added by   such a result from the model checker will be minimal.                                                                                     & Nominal and Ordinal                             & LS6                                                                        \\ \hline
		PRQ4                                 & Such a result from   the model checker could save me time.                                                                                                     & Nominal and Ordinal                             & LS6                                                                        \\ \hline \rowcolor{gray!10}
		\multicolumn{4}{l}{\textbf{Understanding our   Counterexample Explanation Approach Results (Posttest)}}                                                                                                                                                                                                                \\ \hline

		POQ1                                 & The value added by an   approach like counterexample explanation is minimal.                                                                                   & Nominal and Ordinal                             & LS6                                                                        \\ \hline
		POQ2                                 & An approach like   counterexample explanation saves me time.                                                                                                   & Nominal and Ordinal                             & LS6                                                                        \\ \hline
		POQ3                                 & An approach like   counterexample explanation allows me to better understand inconsistencies.                                                                  & Nominal and Ordinal                             & LS6                                                                        \\ \hline
		POQ4                                 & An approach like   counterexample explanation makes me more confident that I really understand   the inconsistencies that I am investigating.                   & Nominal and Ordinal                             & LS6                                                                        \\ \hline \rowcolor{gray!10}
		\multicolumn{4}{l}{\textbf{Counterexample   Explanation Features (Ratings)}}                                                                                                                                                                                                                                          \\ \hline
		FQ1                                  & Translation of   specifications from formal temporal format to natural language-like format.                                                                   & Nominal and Ordinal                             & LS7                                                                        \\ \hline
		FQ2                                  & Listing inconsistent   specification.                                                                                                                          & Nominal and Ordinal                             & LS7                                                                        \\ \hline
		FQ3                                  & Highlighting   sub-parts of the inconsistent specifications that leads to an inconsistency.                                                                       & Nominal and Ordinal                             & LS7                                                                        \\ \hline
		FQ4                                  & Providing the component   name that belongs to the inconsistent specifications.                                                                               & Nominal and Ordinal                             & LS7                                                                        \\ \hline
		FQ5                                  & Providing an expected  nominal behavior in the explanation for the corresponding erroneous states and variables of the counterexample.                                   & Nominal and Ordinal                             & LS7                                                                        \\ \hline
		FQ6                                  & Highlighting the erroneous states and variables in the counterexample.                                                                                           & Nominal and Ordinal                             & LS7                                                                        \\ \hline
		\rowcolor{gray!10}
		\multicolumn{4}{l}{\textbf{Feedbacks (After   completion of the experiment)}}                                                                                                                                                                                                                                              \\ \hline
		FE1                                  & Are inconsistencies easier to understand with the results  created by the counterexample explanation approach in comparison to those of the original model checker? & Nominal and Ordinal                 & LS5                                                                        \\ \hline
		FE2                                  & What challenges did you face while analyzing inconsistencies in a specification with the proposed approach?               & Nominal                             & Not Applicable                                                             \\ \hline
		FE3                                  & Do you think it is easy to maintain consistency with the proposed counterexample explanation while refining requirements into requirements for sub-components? 
  & Nominal and Ordinal                 & LS3                                                                        \\ \hline
		FE4                                  & Do you think the  proposed counterexample explanation approach is usable in real-world development   processes?                                                    & Nominal and Ordinal                 & LS5                                                                        \\ \hline
		FE5                                  & Would  you consider using formal methods with our approach in real-world projects?                          & Nominal and Ordinal                 & LS5                                                                        \\ \hline
		FE6                                  & Would   you consider using the presented approach in your project? If so, please name the project and a contact person?                                                  & Nominal and Ordinal                 & LS5                                                                        \\ \hline
		FE7                                  & Do you think presenting a list of possible suggestions/fixes would be helpful to understand and   fix inconsistencies?                                                    & Nominal and Ordinal                 & LS5                                                                        \\ \hline
		FE8                                  & Suggestions   for further improvements.                                                                                                                        & Nominal                             & Not Applicable                                                             \\ \hline
	\end{tabular}
\end{table*}

\subsection{Tools used for the Study}
\label{sec:survey_tool}

Since we do not require time recording, we will use Microsoft Forms for Excel\footnote{\url{https://support.microsoft.com/en-us/office/surveys-in-excel-hosted-on-the-web-5fafd054-19f8-474c-97ec-b606fcda0ff9}} for the user study that provides required features for performing a survey. Further, it is easily accessible within the company and already familiar to the participants. 
Later, we plan to transfer the results to an Microsoft Excel to perform the analysis.

All content-wise explanations for both \emph{Part\,1} and \emph{Part\,2} of the study are provided as a video and are accessible via an online platform, \eg YouTube or the Bosch-internal equivalent called \emph{BoschTube}.

\section{Participants}
\label{sec:participants}

Our counterexample explanation approach focuses on enhancing safety analysis for automotive systems~\cite{KaleeswaranNVG20}. Thus, we are interested in performing this user study only with Bosch automotive engineers, particularly engineers working on system development, requirement elicitation, and safety analysis. The target population for our study is very specific and thus, it is hard to make a finite list of participants by applying probabilistic sampling. As per Kitchenham and Pfleeger~\cite{KitchenhamP08}, when a target population is very specific and limited, non-probabilistic sampling can be used to identify the participants. Therefore, we intend to use two non-probabilistic sampling methods for \emph{Part\,1} of our study, namely, \emph{convenience sampling} and \emph{snowball sampling}. Further, we invite participants with knowledge on formal methods for \emph{Part\,2} of our study by filtering the participants of \emph{Part\,1} based on the responses to the demographic questions Q1 to Q3.

First, we start with the convenience sampling for \emph{Part\,1}. We send e-mails with the survey link to participants collected through department mailing lists and community mailing lists of all relevant Bosch business units. We perform snowball sampling with the accepted participants by asking for further potential participants at the end of the survey. In the e-mail invitation, we will explicitly mention that the anonymity of results will be preserved. So, while publishing results or sharing the survey responses for evaluation, we will remove all personal, product- and project-related information.

\section{Execution Plan}
\label{sec:execution}

In this section, we describe the execution plan of both the \emph{online survey (Part\,1)} and the \emph{one-group pretest-posttest user study (Part\,2)}, depicted in \fig{steps}.

\subsection{Execution Plan of Part\,1}
\label{sec:plan_part1}

With the accepted participants from the sampling process described in \sect{participants}, we perform an \emph{online survey (Part\,1)} that comprises four steps (\cf \fig{steps}).

First, we notify participants regarding the data processing agreement. Additionally, we also state explicitly that their names, project- and product-related information will be removed while results are shared for evaluation. Then we show a video, welcoming the participant and explaining the background and motivation of this survey. Then, we ask participant to answer the demographic questions (Q1 to Q5 in \tab{survey_questions}), and further the main survey questions (Q6 to Q18 in \tab{survey_questions}). Finally we conclude the survey with a thanks note.

\subsection{Execution Plan of Part\,2}
\label{sec:plan_part2}

For the one-group pretest-posttest user study, we invite participants from \emph{Part\,1} who indicated knowledge of formal methods. Similar to \emph{Part\,1}, \emph{Part\,2} starts with a data processing agreement, followed by a background and motivation video. Our one-group pretest-posttest user study is executed with the invited participants as follows: a pretest experiment, then intervention, and finally the posttest experiment.

\paragraph{Pretest}
The pretest experiment starts with a video demonstrating the pretest experiment with a simple example of an OR-gate behavior. After that, another video explains the system model and specification of an airbag system that serves as a use case for the pretest experiment.

During the actual experiment, the participant analyzes the violated specification and the counterexample returned by the model checker to understand the inconsistent parts of the specification. Further, based on her understanding, the participant answers the task questions (TQ1 to TQ9 except of TQ4 in \tab{onegroup}).
Finally, the pretest is concluded by answering the pre-questionnaire survey questions PRQ1 to PRQ4.

\paragraph{Intervention}
After the pretest experiment, a video explains the counterexample explanation approach~\cite{KaleeswaranNVG20}. This serves as an intervention in our study.   

\paragraph{Posttest}
Like the steps followed for the pretest experiment, the posttest experiment starts with a demonstration video with the same use case of the OR-gate behavior, but this time with the counterexample explanation approach. This is followed by a video that explains the system model and specification of the electronic power steering system (EPS), a commercial Bosch product. Then the participants interpret the explanation provided by the counterexample explanation approach to understand the inconsistency. Based on the explanation,  participants answer the task questions (TQ1 to TQ9 except of TQ3 in \tab{onegroup}). Subsequently, they answer the post-questionnaire survey questions POQ1 to POQ4.

After completing the posttest experiment, participants rate the features (FQ1 to FQ6 in \tab{onegroup}) provided by the counterexample explanation approach and respond to the feedback questions (FE1 to FE8 in \tab{onegroup}). Finally, \emph{Part\,2} of our study concludes with a thanks note to the participants.

\section{Analysis Plan}
\label{sec:analysis}

To obtain the results from the study, we follow the recommendation by Robbins and Heiberger~\cite{Robbins2011}. To plot the demographic questions (Q1 to Q5), we use a normal bar chart. The graph we intend to use is a diverging stacked bar chart with counts (see  Figure~10 in \cite{Robbins2011}) to plot the results for questions (Q6 to Q18, FQ1 to FQ6, and FE1 to FE8). The X-axis label of the graph shows counts and percentages, the Y-axis label shows the demographic answers. To present the pretest and posttest experiment results in a comparative way, we use grouped bar charts. For the comparative graph of task questions (TQ1 to TQ9), X-axis labels are individual questions, and Y-axis are the number of correct answers. Likewise, for understanding the result questions (PRQ1 to PRQ4 and POQ1 to POQ4), X-axis labels are scale values (\cf \tab{likert_scale}), and Y-axis is the count for every scale value. We do not associate demographic answers for plotting the comparative graphs. To make a reliable argument, we use the demographic answers for discussing the comparative graph. For example: \emph{"2 out of 10 participants who have more than ten years of experience answered TQx correctly"}.

Qualitative statements received from participants are gathered, organized, and summarized individually for every question. We summarize the qualitative statements through the following three steps: \textbf{(i)~Microanalysis:} The answers from participants are gone through individually by the first author and he assigns labels to statements. The rest of the authors will validate the initial labels and provide feedback for improvement. At the end of this step, all authors come to a mutual agreement on the initial labels. \textbf{(ii)~Categorization:} Based on the feedback for improvisation, the first author performs second iteration. As a result, a set of themes are extracted which are deemed to be essential. \textbf{(iii)~Saturation:} This is the final step where all the authors come to the final agreement on labels, themes, and summarized statements. Since the qualitative statement is a medium to express an individual opinion, the categorization of labels are associated with the demographic answers. For example: \emph{"an engineer who has seven years of experience states that the counterexample explanation approach can promote the usage of model checkers among system engineers"}.
\section{Threats to Validity}
\label{sec:tov}

In this section, we discuss threats that may jeopardize the validity of our study results as well as measures we take to reduce these threats.

We consider threats to validity as discussed by Wohlin \etal~\cite{WohlinRHO12}, Kitchenham and Pfleeger~\cite{KitchenhamP08}, and Campbell and Stanley~\cite{CampbellS63}. In the following, we structure them according to construct validity, internal validity, and external validity.

\paragraph{Construct Validity}

The prime threats to construct validity are related to the completeness of the questionnaire and in phrasing questions in a way that is understood by all participant in the same way. To mitigate these threats, we have taken the following measures during our survey preparation: (i)~we incorporated feedback from two senior engineers with a background in formal methods and model checking, (ii)~we incorporated feedback regarding unbiased questions from a psychologist, and (iii)~we intend to perform a pilot test with five research engineers to check for completeness and understandability.

\paragraph{Internal Validity}

The critical internal threat to be considered for the online survey is the selection of participants. Since we follow snowball sampling for participant selection, there could be a possibility of several participants working in the same project, which could bias the final result. Therefore, we will consider only a small number of participants from each project and neglect further project members.

We consider threats to internal validity listed by Campbell and Stanley~\cite{CampbellS63} for the pretest-posttest experiment. To mitigate the \emph{history} and \emph{maturation} threats, we plan to perform both the pretest and the posttest experiments on the same day. The most severe threats to be considered in this experimental design are \emph{testing} and \emph{instrumentation}. Those threats arise because participants get overwhelmed with the intervention. Consequently, participants could answer more positively in the posttest experiment than the actual value. 

To mitigate these threats, we state to the participants explicitly that the obtained study results will serve as a reference in the future to use our counterexample explanation approach for real-world projects at Bosch.
Additionally, to avoid overwhelmed responses and accept only valid responses, we will cross-check the answers provided for the task questions (TQ1 to TQ9). Further, to reduce biasing between the pretest and the posttest experiment, the use case of an airbag system (a toy example) used in the pretest is significantly less complex than the use case of the Bosch EPS system. However, to adjust the difficulty level of the systems used for the experiment, we plan to perform a pilot study with five research engineers. Adjustment of difficulty will be done by increasing or decreasing the number of components and size of the specifications that need to be understood by the participants.

\paragraph{External Validity}

One of the severe drawbacks of the one-group pretest-posttest experiment is its generalization. However, the benefit of our study is that we use a real-world system for the posttest experiment, and the participants are professional engineers who work on real-world automotive projects at Bosch.

\section{Implications}
\label{sec:mmplications}

\emph{Part\,1} of our study will find the importance and difficulty of finding inconsistent specifications introduced during the refinement of top-level specifications as well as the necessities, acceptance, and challenges of using formal methods at Bosch. Further with \emph{Part\,2} of our study, we will evaluate whether our counterexample explanation approach is beneficial for the difficulties and challenges identified in \emph{Part\,1} and therefore for the adoption of formal methods in industrial projects at Bosch.

\bibliographystyle{IEEEtran}
\bibliography{ICSME2021}

\end{document}